\documentclass[cameraready]{Interspeech}
\usepackage{subfigure}
\usepackage{booktabs}
\usepackage{makecell}
\usepackage[table]{xcolor} 
\usepackage{colortbl}      
\usepackage{multirow}
\usepackage{makecell}
\usepackage{tabularx}
\usepackage{cite}
\definecolor{lightgray}{gray}{0.92}
\title{BiEAR: A Human Auditory-Inspired Adaptive Binaural Front-end for Multi-Speaker Localisation and Distance Estimation}
\author[affiliation={1}, orcid=0009-0004-6123-0587]{Hanyu}{Meng}
\author[affiliation={1}, orcid=0000-0003-4673-6534]{Eliathamby}{Ambikairajah}
\author[affiliation={1}, orcid=0000-0001-8492-1787]{Vidhyasaharan}{Sethu}
\author[affiliation={2}, orcid=0009-0006-5795-2899]{Qiquan}{Zhang}
\author[affiliation={3}, orcid=0000-0001-9158-9401]{Haizhou}{Li}
\address{
    $^1$ The University of New South Wales, Sydney, Australia \\
    $^2$ Tongyi Speech Lab, Alibaba Group, Hangzhou, China \\
    $^3$ School of Artificial Intelligence, The Chinese University of Hong Kong, Shenzhen, China
}
\email{hanyu.meng@unsw.edu.au, e.ambikairajah@unsw.edu.au}

\usepackage{comment}
\begin{document}
\maketitle
\vspace{-2mm}
\begin{abstract}
\vspace{-1mm}
We present BiEAR, a human auditory-inspired adaptive binaural front-end for multi-speaker localisation and distance estimation. Inspired by medial olivocochlear (MOC) feedback in human hearing, BiEAR uses a neural controller to adaptively adjust the frequency selectivity of a binaural auditory filterbank during inference. This yields time–frequency adaptive representations for ears, enabling the model to respond to changing acoustic conditions. We evaluate BiEAR on multi-speaker localisation and distance estimation in anechoic and real-room environments. Results show that the adaptive front-end improves localisation accuracy and robustness to unseen speakers and rooms compared with commonly used fixed binaural front-ends. Visualisation and analysis of learned filter adaptations show that BiEAR emphasises informative frequency bands over time. These findings suggest that adaptive, biologically inspired binaural front-ends can improve machine hearing robustness in complex acoustic scenes\footnote{Code: \url{https://github.com/Hanyu-Meng/BiEAR}.}. 


\end{abstract}
\vspace{-1mm}

\vspace{-1mm}
\section{Introduction}
\vspace{-1mm}
Binaural sound source localisation and distance detection support many machine-hearing applications, including binaural speech enhancement~\cite{binaural_SE_1,binaural_SE_2}, acoustic scene analysis~\cite{ASA_1,ASA_2, lee2025deepasa}, and speaker tracking for robotics~\cite{track_1, track_2, ipdnet}. In computational auditory scene analysis (CASA)~\cite{CASA}, achieving human-like localisation and distance perception remains challenging in complex and dynamically changing acoustic environments~\cite{spille14_interspeech}.

To address this, recent deep learning-based auditory-inspired binaural localisation models integrate multiple binaural cues, including auditory-scaled spectrograms~\cite{Vecchiotti2019WaveLoc, fu25_interspeech, Kuang2025BASTMamba}, interaural phase/level differences~\cite{Wang2023FramewiseMSSL, DeepEar, Jazaeri2025MultiSpeakerDOA, auditory_inspired_binaural}, generalised cross-correlation~\cite{Tokala2025BinauralLocEAA}, and other inter-channel relative features~\cite{Yang2021DeepDPRTF}. However, most adopt fixed inference graphs, which limits adaptation to non-stationary scenes and unseen environments. Moreover, these front-ends are largely feedforward and omit adaptive efferent feedback, a key feature of human hearing~\cite{Guinan2018Olivocochlear}.

\begin{figure}[!t]
  \centering
  \vspace{-0.2em}
  \includegraphics[width=0.45\textwidth]{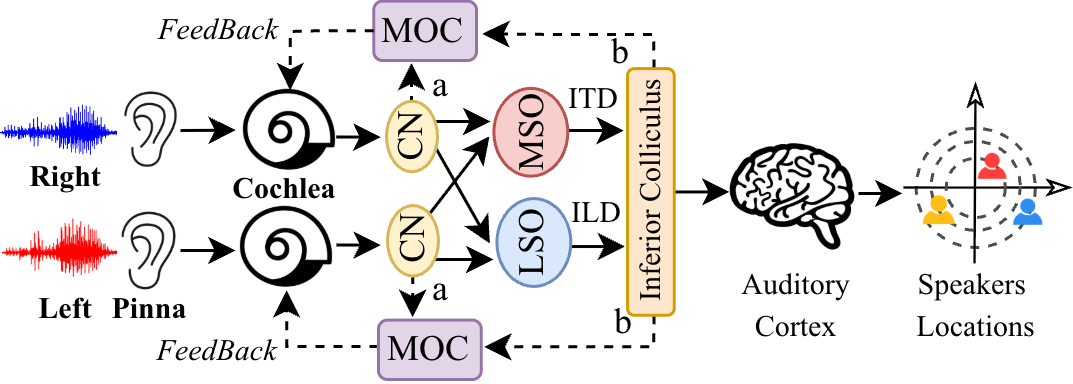}
  \vspace{-0.8em}
  \caption{Overview of the human binaural auditory system with MSO/LSO-based ITD/ILD extraction and adaptive MOC efferent modulation.}
  \label{fig:intro}
  \vspace{-2.4em}
\end{figure}

As illustrated in Fig.~\ref{fig:intro}, incoming binaural signals are first shaped by the pinna and decomposed in the cochlea, before being relayed via the cochlear nucleus (CN) to brainstem nuclei for spatial cue extraction. Specifically, the medial superior olive (MSO) computes interaural time differences (ITDs), while the lateral superior olive (LSO) encodes interaural level differences (ILDs). These binaural cues are subsequently integrated in higher auditory centres, such as the inferior colliculus (IC) and the auditory cortex, to support spatial hearing~\cite{Grothe2010SoundLocalization,Ashida2011Jeffress}. Crucially, this predominantly feedforward pathway is modulated by the medial olivocochlear (MOC) efferent system, which dynamically regulates cochlear gain and frequency selectivity tuning via outer hair cells through both fast CN-driven reflexive loops (path~a) and top-down feedback from higher auditory centres (path~b)~\cite{Guinan2018Olivocochlear,Andeol2011AuditoryEfferents}. The latter requires task-specific prior knowledge and is therefore beyond the scope of this paper.

Recent studies have proposed auditory-inspired monaural adaptive audio front-end that simulate cochlear processing with neural feedback controllers, and have shown effectiveness in improving robustness and performance across diverse downstream tasks~\cite{meng2025adaptive,qiquan_AdaFE,buddhi_AdaFE}. However, these models focus exclusively on monaural processing and thus may not fully reflect human binaural hearing functions. Motivated by these limitations, we propose BiEAR, a human-auditory inspired binaural adaptive front-end with bilateral feedback controllers that dynamically regulate auditory filter characteristics based on frame subband sound pressure levels (SPL) during inference. 

Overall, our contributions are: (1) We propose BiEAR, a human auditory-inspired binaural adaptive front-end that incorporates MOC-inspired neural feedback to dynamically regulate filterbank selectivity by controlling subband Q-factors during inference. (2) We design ear-specific neural feedback controllers and two frame-wise Q-modulation strategies (absolute and relative control), enabling time--frequency adaptive and potentially asymmetric filtering in the left and right pathways. (3) We evaluate BiEAR on multi-speaker localisation and distance estimation, demonstrating improved robustness to unseen acoustic environments and providing interpretable analysis via visualisation of the learned filterbank dynamics.

\vspace{-2mm}
\section{BiEAR}\label{sec:method}
\vspace{-2mm}
\begin{figure*}[!ht]
  \centering
  \subfigure[Overview of the proposed BiEAR framework.]{
    \includegraphics[width=0.78\textwidth]{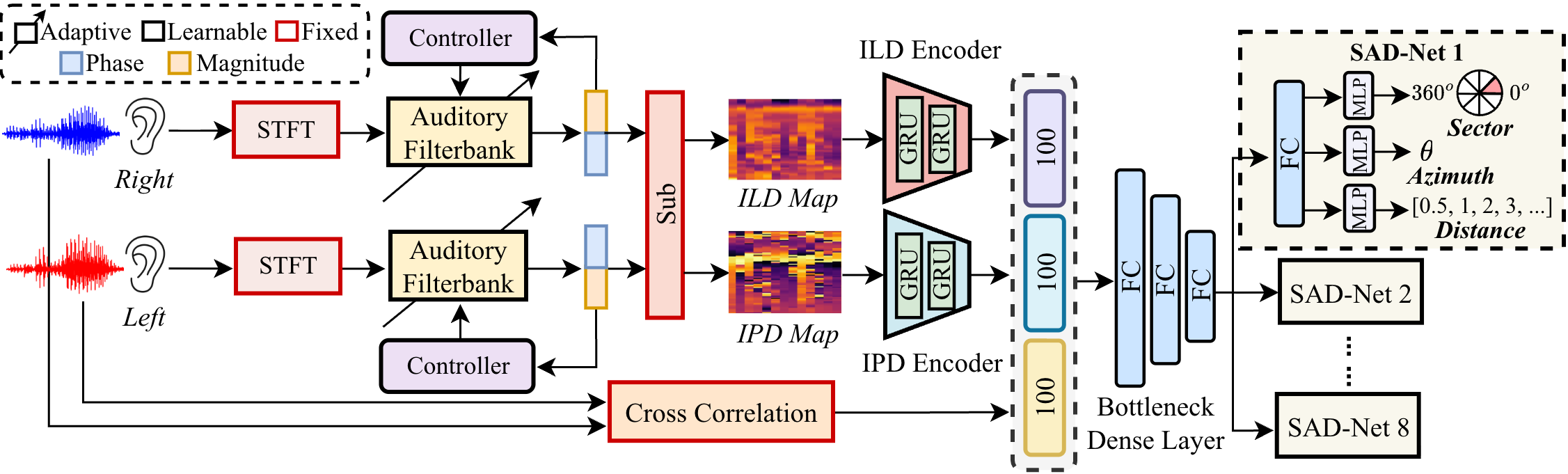}
    \label{fig:biear}
  }
  \hfill
  \subfigure[Controller Structure.]{
    \includegraphics[width=0.18\textwidth]{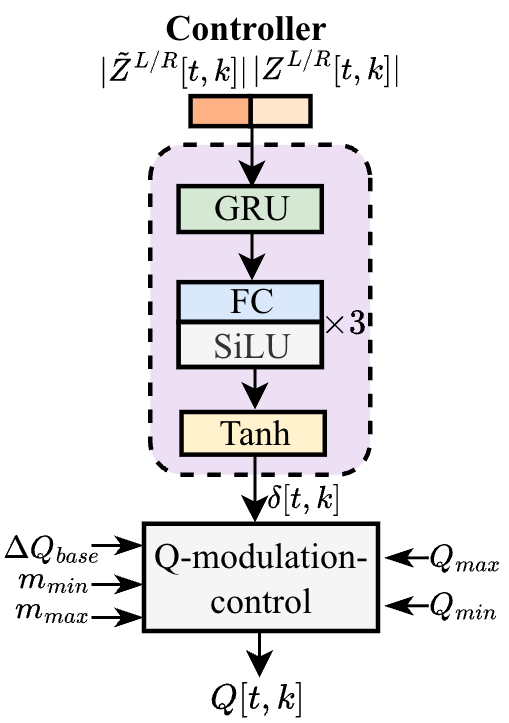}
    \label{fig:neural_controller}
  }
  \vspace{-1em}
  \caption{The overview of the proposed binaural model with a feedback-controlled adaptive front-end.}
  \vspace{-1.8em}
  \label{fig:biear_overview}
\end{figure*}
\vspace{-1mm}
\subsection{Model Overview}\label{sec:model_overview}
\vspace{-1mm}
An overview of BiEAR is shown in Fig.~\ref{fig:biear}. Following prior work~\cite{DeepEar,fu25_interspeech}, we uniformly partition the azimuth range $[0^\circ,360^\circ)$ into eight $45^\circ$ sectors. 
Accordingly, BiEAR comprises eight sector-wise SAD-Nets for joint source detection, azimuth estimation, and distance classification. Given the binaural input, each SAD-Net jointly predicts: (i) whether an active source is present in the sector, (ii) the source azimuth if a source is detected, and (iii) the distance class of each detected sound source.
\vspace{-3mm}
\subsection{Adaptive Binaural Features}
\vspace{-1mm}
Given a one-second binaural waveform segment $x^{L/R}[n]$, we perform framing and short-time Fourier transform (STFT) with a Hann window, obtaining the complex time--frequency representation $X^{L/R}[t,f]$.

To form an auditory-inspired subband representation, we group the STFT coefficients using $K$ adjustable Gabor filters~\cite{Zeghidour2021LEAF} distributed on the Equivalent Rectangular Bandwidth (ERB) scale~\cite{erb_scale}, the filtering process in frequency domain can be expressed as:
\vspace{-0.5mm}
\begin{equation}
\vspace{-1mm}
\begin{aligned}
Z^{L/R}[t,k] &= \sum_f W_t^k(f) X^{L/R}[t,f], \\
W_t^k(f) &= \exp\left(-\frac{(f-f_c^k)^2}{2(\mathrm{BW}_t^k)^2}\right).
\end{aligned}
\vspace{-1mm}
\end{equation}
where $k \in \{0,1,\ldots,K-1\}$ is the subband index, and $f_c^k$ and $\mathrm{BW}_t^k$ denote the centre frequency and time-varying adaptive bandwidth of the $k$-th filter, respectively. The grouped representation $Z^{L/R}[t,k] \in \mathbb{C}$ preserves both magnitude and phase, which are required for binaural cue extraction.

The neural controller modulates frequency selectivity by adjusting the filter quality factor (Q-factor), defined as $Q_t^k = f_c^k / \mathrm{BW}_t^k$. Following prior monaural adaptive front-end~\cite{qiquan_AdaFE,buddhi_AdaFE}, Q control effectively regulates subband selectivity. With fixed $f_c^k$, increasing $Q_t^k$ narrows $\mathrm{BW}_t^k$ and increases filter gain, yielding a more selective response in the corresponding subband, analogous to frequency dependent auditory attention. Details of the proposed Q-control mechanisms are provided in Section~\ref{sec:neural_control}. From $Z^{L/R}[t,k]$, we derive binaural spatial cues, including interaural level difference (ILD) and interaural phase difference (IPD), as
\vspace{-1mm}
\begin{equation}
\vspace{-1mm}
\mathrm{ILD}[t,k]=20 \log_{10} \frac{|Z^{L}[t,k]| + \varepsilon}{|Z^{R}[t,k]| + \varepsilon},
\end{equation}
\vspace{-1.3mm}
\begin{equation}
\mathrm{IPD}[t,k] = \operatorname{arg}\bigl(
Z^{L}[t,k]\,(Z^{R}[t,k])^{*} \bigr),
\end{equation}
where $\varepsilon$ is a small constant for numerical stability.

In BiEAR, we compute binaural cross-correlation (CC) directly from the raw waveform as an auditory-inspired coarse cue of interaural similarity and broadband signal level~\cite{Lindemann1986}, and use it together with the ILD and IPD maps as inputs to the back-end networks. As shown in Fig.~\ref{fig:biear}, we compress the temporal dimension of the ILD and IPD maps using two GRU layers (hidden dimensions: 200 and 100), yielding two 100-dimensional embeddings. For dimensional consistency, CC is computed over $\pm 3$\,ms interaural delays, producing a 100-dimensional representation.
\vspace{-2mm}
\subsection{Q-Factor Modulation and Control}
\label{sec:neural_control}
\vspace{-1mm}
\subsubsection{Neural Feedback Controller}
\vspace{-1mm}
We design a neural feedback controller that emulates auditory efferent feedback and dynamically adjusts filter selectivity based on the input. As shown in Fig.~\ref{fig:neural_controller}, each controller comprises a GRU layer (hidden size 128) followed by three Sigmoid Linear Unit (SiLU) activated fully connected (FC) layers with output sizes 128, 128, and $K$, respectively, and the final layer is hyperbolic tangent nonlinearity, yielding a bounded control signal $\delta[t,k] \in [-1,1]$.
 
At each frame $t$, the controller input concatenates the instantaneous subband SPL $E^{L/R}[t,k]=|Z^{L/R}[t,k]|$ and a smoothed SPL trace $\tilde{E}^{L/R}[t,k]$ computed by an exponential moving average:
$\tilde{E}[t,k]=\beta\,\tilde{E}[t-1,k]+(1-\beta)\,E[t,k]$, where $\beta\in[0,1]$ controls the temporal memory. The controller outputs a frame-wise, frequency dependent modulation signal that adjusts the filterbank Q-factors for the current frame. The baseline Q-factor $Q_0[k]$ is initialised from a standard ERB-spaced filterbank with $K$ subbands.
\vspace{-2.5mm}
\subsubsection{Frequency Dependent Q Variation Profile}
\vspace{-1mm}
Following prior work~\cite{buddhi_AdaFE,qiquan_AdaFE}, we incorporate frequency dependent flexibility into the Q modulation by defining a base Q-variation profile whose range increases with centre frequency. We first map each centre frequency to the ERB-rate scale,
$e_k=\mathrm{ERBRate}(f_c^k)$, and normalise it using the minimum and maximum ERB-rate values across subbands:
\vspace{-0.3em}
\begin{equation}
\begin{aligned}
u_k &= \frac{e_k - e_{\min}}{e_{\max} - e_{\min}}, \\
\Delta Q_k &= \Delta Q_{\mathrm{base}}
\Big(m_{\mathrm{low}} + (m_{\mathrm{high}} - m_{\mathrm{low}})\,u_k\Big),
\end{aligned}
\vspace{-2mm}
\end{equation}
where $u_k\in[0,1]$ is the normalised ERB rate, and $m_{\mathrm{low}}$, $m_{\mathrm{high}}$, and $\Delta Q_{\mathrm{base}}$ control the minimum, maximum, and overall magnitude of the Q-variation range. Consequently,
$\Delta Q_k \in[\Delta Q_{\mathrm{base}}m_{\mathrm{low}},\Delta Q_{\mathrm{base}}m_{\mathrm{high}}]$.
\vspace{-1.5mm}
\subsubsection{Q-Factor Control Strategies}
\vspace{-1mm}
We combine the baseline Q-factor $Q_0[k]$, the frequency dependent variation range $\Delta Q[k]$, and the controller output $\delta[t,k]$ via two frame-wise control strategies: \textbf{absolute} and \textbf{relative} Q control.

\textbf{Absolute Q control:} The controller applies an additive offset to the baseline Q-factor:
\vspace{-0.5em}
\begin{equation}
Q^{\mathrm{abs}}[t,k]= \mathrm{clip}\left(Q_0[k] + \Delta Q[k]\, \delta[t,k],
Q_{\min}, Q_{\max}\right),
\end{equation}
where $\mathrm{clip}(\cdot)$ bounds the Q-factor within $[Q_{\min},Q_{\max}]$ for numerical stability.

\textbf{Relative Q control:} Alternatively, the controller modulates selectivity multiplicatively by scaling the baseline Q-factor:
\vspace{-0.5em}
\begin{equation}
Q^{\mathrm{rel}}[t,k]=\mathrm{clip}\left(Q_0[k]\big(1 + \Delta Q[k]\, \delta[t,k]\big),
Q_{\min}, Q_{\max}\right).
\end{equation}

Absolute control applies an additive offset to $Q_0[k]$, resulting in a uniform additive modulation across subbands, whereas relative control scales $Q_0[k]$ multiplicatively, producing Q changes proportional to the Q value baseline and thus inherently frequency dependent.

\vspace{-2mm}
\subsection{Back-end and Training Approach}
\vspace{-1mm}
The back-end network takes as input the concatenation of three 100-dimensional binaural feature embeddings. This 300-dimensional vector is first projected through a bottleneck of three FC layers with output sizes of 512, 400, and 200.

The bottleneck output is then fed into $S=8$ sector-wise SAD-Nets (Section~\ref{sec:model_overview}). Each SAD-Net starts with a shared FC layer of 100 units and branches into three task-specific MLP heads for source detection, azimuth estimation, and distance classification. Each head comprises three FC layers with output sizes 50, 10, and $1$ for source detection and azimuth regression or the number of distance classes for distance classification. All FC layers use Rectified Linear Unit (ReLU) activations.

We train the model with a joint multi-task objective:
\vspace{-0.8em}
\begin{equation}
\begin{aligned}
\mathcal{L}_{\text{total}}
= \sum_{s=1}^{S} \Big[
&\lambda_{1}\mathcal{L}_{\text{BCE}}(y_s,\hat{y}_s) \\
&+ \mathcal{I}_s \Big(
\lambda_{2}\mathcal{L}_{\text{MSE}}(\theta_s,\hat{\theta}_s)
+ \lambda_{3}\mathcal{L}_{\text{CE}}(\mathbf{d}_s,\hat{\mathbf{d}}_s)
\Big)
\Big],
\end{aligned}
\label{eq:loss}
\end{equation}
where $\lambda_{1}$--$\lambda_{3}$ balance the three tasks. Here, $y_s$ and $\hat{y}_s$ are the ground truth and predicted source presence logits for sector $s$. The indicator $\mathcal{I}_s\in\{0,1\}$ masks the azimuth and distance losses so they are applied only when a source is active in sector $s$. Finally, $\theta_s$ and $\hat{\theta}_s$ denote the ground truth and estimated azimuth angles, and $\mathbf{d}_s$ and $\hat{\mathbf{d}}_s$ are the ground-truth and predicted categorical distance distributions.
\vspace{-2mm}
\section{Experimental Setups}\label{sec:experiments}
\vspace{-0.5mm}
\subsection{Data Preparation}\label{sec:experiments-data}
\vspace{-1mm}
An anechoic binaural speech dataset was generated following the DeepEar protocol~\cite{DeepEar}. Clean monaural utterances were drawn from TIMIT~\cite{timit}, clipped or zero-padded to 1 second, and spatialized by convolving with anechoic binaural room impulse responses (BRIRs). Multi-speaker mixtures were created by independently spatializing multiple utterances with randomly sampled azimuths and distances and then summing the resulting binaural signals. All BRIR datasets used for anechoic training and reverberant evaluation are sourced from the TU Berlin database~\cite{wierstorf2011free} and are summarized in Table~\ref{tab:brir}. The resulting anechoic training set (\textit{Anechoic-train}) contains 72{,}000 samples, while the validation and test sets (\textit{Anechoic-val} and \textit{Anechoic-test}) each contain 9{,}000 samples, with equal proportions of one-, two-, and three-speaker mixtures. We further construct a speaker-independent test set (\textit{Anechoic-test-unseen-spk}) of 9{,}000 samples using TIMIT speakers disjoint from training.

To evaluate robustness and adaptation in practical environments, we use BRIR recordings from two real rooms: a low-reverberation meeting room and a highly reverberant lecture hall as shown in Table~\ref{tab:brir}. For each room, we generate a 9{,}000-sample test set using utterances from the TIMIT TEST split, ensuring all speakers are unseen. We report performance both without adaptation and with environment transfer, where the anechoic pre-trained model is fine-tuned on 10\% of the reverberant test data and evaluated on the remaining 90\%.

Because the reverberant BRIR distances do not exactly match the discrete distance classes used during anechoic training, distance estimation is formulated with five classes: four in-range classes (0.5~m, 1~m, 2~m, and 3~m) and an additional \textit{other} class for distances $>3$~m. For samples within 3~m, predictions are counted as correct if mapped to the nearest training distance class.
\begin{table}[!t]
\centering
\fontsize{8pt}{7pt}\selectfont
\vspace{-0.8em}
\caption{Summary of BRIR datasets and the corresponding azimuth and distance ranges used for training and evaluation.}
\label{tab:room_config}
\vspace{-0.8em}
\setlength{\tabcolsep}{2pt}
\begin{tabular}{@{}c c c c@{}}
\toprule
Environment & BRIR & Azimuth ($^{\circ}$) & Distance (m) \\
\midrule
Anechoic
& \makecell[c]{\mbox{Anechoic}~\cite{anechoic}}
& 0--360
& 0.5/1/2/3  \\
Meeting Room
& \makecell[c]{\mbox{Spirit}~\cite{spirit}}
& 90--270
& 2  \\
Lecture Hall
& \makecell[c]{\mbox{Auditorium3}~\cite{auditorium3}}
& 90--270
& 1.5/2.93/3.97/5.49  \\
\bottomrule
\end{tabular}
\vspace{-5.5mm}
\label{tab:brir}
\end{table}
\vspace{-0.5mm}
\begin{table*}[!ht]
\setlength{\belowcaptionskip}{-0.5cm}
\setlength{\tabcolsep}{4pt}
\caption{Performance Comparison in Anechoic (Seen / Unseen Speakers; ``w/o'' denotes without)}
\vspace{-0.8em}
\label{tab:performance}
\centering
\fontsize{10pt}{10pt}\selectfont
\resizebox{1\textwidth}{!}{%
\begin{tabular}{l|c|ccccccccc}
\toprule
\multicolumn{1}{c|}{\textbf{Source Type}}
& \multirow{2}{*}{\centering \textbf{\#Params}}
& \multicolumn{3}{c}{\textbf{1-speaker}}
& \multicolumn{3}{c}{\textbf{2-speakers}}
& \multicolumn{3}{c}{\textbf{3-speakers}} \\
\cmidrule(lr){3-5}\cmidrule(lr){6-8}\cmidrule(lr){9-11}
\multicolumn{1}{c|}{\textbf{Metrics}}
& 
& \textbf{Sound Detect Acc (\%)}
& \textbf{Azim. MAE ($^\circ$)}
& \textbf{Dist. Acc (\%)}
& \textbf{Sound Detect Acc (\%)}
& \textbf{Azim. MAE ($^\circ$)}
& \textbf{Dist. Acc (\%)}
& \textbf{Sound Detect Acc (\%)}
& \textbf{Azim. MAE ($^\circ$)}
& \textbf{Dist. Acc (\%)} \\
\midrule
DeepEar \cite{DeepEar}
& \textit{2.08 M}
& 99.78 \,/\, 99.78 & 0.80 \,/\, 0.82 & 95.03 \,/\, 95.13
& 95.19 \,/\, 95.19 & 5.09 \,/\, 5.73 & 85.42 \,/\, 83.39
& 89.23 \,/\, 88.62 & 10.27 \,/\, 10.40 & 73.07 \,/\, 72.01 \\
AuralNet \cite{fu25_interspeech}
& \textit{1.37 M}
& 99.58 \,/\, 99.50 & 0.73 \,/\, 0.78 & \textbf{98.12 \,/\, 97.89}
& 96.00 \,/\, 95.94 & 3.82 \,/\, 3.83 & \textbf{89.12 \,/\, 88.78}
& 89.80 \,/\, 88.90 & 9.23 \,/\, 9.45 & \textbf{75.45 \,/\, 74.78} \\
BiEAR w/o Controller
& \textit{1.29 M}
& 99.62 \,/\, 99.64 & 0.63 \,/\, 0.61 & 96.73 \,/\, 96.95
& 94.07 \,/\, 94.20 & 4.67 \,/\, 4.64 & 82.45 \,/\, 82.42
& 86.91 \,/\, 86.16 & 10.47 \,/\, 10.62 & 69.00 \,/\, 68.55 \\
BiEAR + Single Controller + Abs.
& \textit{1.54 M}
& 99.65 \,/\, 99.61 & 0.54 \,/\, 0.58 & 96.99 \,/\, 96.97
& 94.02 \,/\, 94.12 & 4.51 \,/\, 4.48 & 82.40 \,/\, 82.52
& 86.47 \,/\, 86.59 & 10.26 \,/\, 10.37 & 68.92 \,/\, 68.73 \\
BiEAR + Single Controller + Rel.
& \textit{1.54 M}
& 99.62 \,/\, 99.65 & 0.57 \,/\, 0.53 & 97.00 \,/\, 97.00
& 94.22 \,/\, 94.23 & 4.40 \,/\, 4.47 & 82.93 \,/\, 82.76
& 86.73 \,/\, 86.18 & 10.19 \,/\, 10.39 & 69.46 \,/\, 69.04 \\
BiEAR + Dual Controller + Abs.
& \textit{1.63 M}
& 99.85 \,/\, 99.76 & 0.43 \,/\, 0.48 & 97.65 \,/\, 97.54
& 96.50 \,/\, 96.25 & 3.28 \,/\, 3.47 & 85.72 \,/\, 85.42
& 90.30 \,/\, 89.55 & 8.43 \,/\, 8.71 & 72.75 \,/\, 72.16 \\
BiEAR + Dual Controller + Rel.
& \textit{1.63 M}
& \textbf{99.90 \,/\, 99.80} & \textbf{0.36 \,/\, 0.39} & 97.84 \,/\, 97.65
& \textbf{96.85 \,/\, 96.77} & \textbf{3.05 \,/\, 3.13} & 86.61 \,/\, 86.66
& \textbf{90.82 \,/\, 90.72} & \textbf{8.03 \,/\, 8.18} & 73.91 \,/\, 73.65 \\
\bottomrule
\end{tabular}%
}
\vspace{-3mm}
\end{table*}
\begin{table*}[!ht]
\caption{Performance comparison in practical environments (all unseen speakers \& different acoustic conditions).}
\vspace{-0.8em}
\label{tab:performance_rooms}
\centering
\fontsize{10pt}{10pt}\selectfont
\setlength{\tabcolsep}{4pt}
\resizebox{1\textwidth}{!}{%
\begin{tabular}{c|c|ccc|ccc|ccc}
\toprule
\multirow{2}{*}{\textbf{Room}}
& \multirow{2}{*}{\makecell[c]{\textbf{Source Type}\\[3pt]\textbf{Metrics}}}
& \multicolumn{3}{c|}{\textbf{1-speaker}}
& \multicolumn{3}{c|}{\textbf{2-speakers}}
& \multicolumn{3}{c}{\textbf{3-speakers}} \\
\cmidrule(lr){2-2}\cmidrule(lr){3-5}\cmidrule(lr){6-8}\cmidrule(lr){9-11}
& &
\textbf{Sound Detect Acc (\%)} & \textbf{Azim. MAE ($^\circ$)} & \textbf{Dist. Acc (\%)}
& \textbf{Sound Detect Acc (\%)} & \textbf{Azim. MAE ($^\circ$)} & \textbf{Dist. Acc (\%)}
& \textbf{Sound Detect Acc (\%)} & \textbf{Azim. MAE ($^\circ$)} & \textbf{Dist. Acc (\%)} \\
\midrule
\multirow{6}{*}{\centering\textbf{Meeting Room}}
& DeepEar \cite{DeepEar}
& 66.64 & 15.19 & 8.05
& 63.00 & 20.86 & 7.06
& 53.45 & 23.53 & 6.82 \\
& \quad + env. transfer
& \cellcolor{lightgray}85.18 & \cellcolor{lightgray}6.30 & \cellcolor{lightgray}83.33
& \cellcolor{lightgray}86.38 & \cellcolor{lightgray}11.27 & \cellcolor{lightgray}85.81
& \cellcolor{lightgray}86.14 & \cellcolor{lightgray}13.28 & \cellcolor{lightgray}85.42 \\
& AuralNet \cite{fu25_interspeech}
& 64.97 & 14.47 & 61.91
& 63.24 & 17.56 & 54.23
& 63.44 & 18.17 & 52.25 \\
& \quad + env. transfer
& \cellcolor{lightgray}88.31 & \cellcolor{lightgray}5.19 & \cellcolor{lightgray}92.19
& \cellcolor{lightgray}91.04 & \cellcolor{lightgray}7.90 & \cellcolor{lightgray}91.90
& \cellcolor{lightgray}93.46 & \cellcolor{lightgray}8.07 & \cellcolor{lightgray}92.46 \\
& BiEAR + Dual Controller + Rel.
& 70.39 & 12.31 & 69.73
& 78.52 & 14.41 & 64.44
& 78.91 & 14.85 & 62.76 \\
& \quad + env. transfer
& \cellcolor{lightgray}\textbf{93.74} & \cellcolor{lightgray}\textbf{3.92} & \cellcolor{lightgray}\textbf{93.98}
& \cellcolor{lightgray}\textbf{93.45} & \cellcolor{lightgray}\textbf{6.87} & \cellcolor{lightgray}\textbf{93.70}
& \cellcolor{lightgray}\textbf{95.51} & \cellcolor{lightgray}\textbf{6.59} & \cellcolor{lightgray}\textbf{95.26} \\
\midrule
\multirow{6}{*}{\centering\textbf{Lecture Hall}}
& DeepEar \cite{DeepEar}
& 72.81 & 14.58 & 11.86
& 63.00 & 19.63 & 17.10
& 53.45 & 24.64 & 22.60 \\
& \quad + env. transfer
& \cellcolor{lightgray}80.88 & \cellcolor{lightgray}7.75 & \cellcolor{lightgray}78.15
& \cellcolor{lightgray}73.78 & \cellcolor{lightgray}13.21 & \cellcolor{lightgray}67.92
& \cellcolor{lightgray}66.13 & \cellcolor{lightgray}18.82 & \cellcolor{lightgray}59.84 \\
& AuralNet \cite{fu25_interspeech}
& 71.03 & 13.99 & 68.64
& 71.99 & 15.64 & 63.84
& 71.59 & 18.26 & 58.66 \\
& \quad + env. transfer
& \cellcolor{lightgray}88.79 & \cellcolor{lightgray}4.27 & \cellcolor{lightgray}90.89
& \cellcolor{lightgray}82.46 & \cellcolor{lightgray}9.15 & \cellcolor{lightgray}80.83
& \cellcolor{lightgray}80.88 & \cellcolor{lightgray}13.71 & \cellcolor{lightgray}71.64 \\
& BiEAR + Dual Controller + Rel.
& 74.92 & 10.87 & 74.72
& 72.69 & 14.45 & 66.19
& 70.72 & 17.74 & 58.98 \\
& \quad + env. transfer
& \cellcolor{lightgray}\textbf{93.22} & \cellcolor{lightgray}\textbf{3.35} & \cellcolor{lightgray}\textbf{90.89}
& \cellcolor{lightgray}\textbf{85.55} & \cellcolor{lightgray}\textbf{8.74} & \cellcolor{lightgray}\textbf{80.84}
& \cellcolor{lightgray}\textbf{82.93} & \cellcolor{lightgray}\textbf{12.73} & \cellcolor{lightgray}\textbf{72.57} \\
\bottomrule
\end{tabular}%
}
\vspace{-0.2em}
\end{table*}
\vspace{-0.5em}
\subsection{Model Configurations}\label{sec:experiments-model}
\vspace{-0.8mm}
We compare BiEAR with AuralNet~\cite{fu25_interspeech}, the current state-of-the-art 3D binaural localisation model, and DeepEar~\cite{DeepEar}, a strong and widely used auditory-inspired baseline. Since these methods share a similar binaural multi-task back-end design with our framework, we ensure a fair comparison by keeping a unified back-end for all methods and varying only the front-end components.

All models were trained with the joint loss in Eq.~\ref{eq:loss} and optimised using Adam~\cite{Adam}, with $\lambda_{1}=0.25$, $\lambda_{2}=0.45$, and $\lambda_{3}=0.35$. Training was conducted for up to 100 epochs with a batch size of 64, early stopping (patience = 10), an initial learning rate of $10^{-4}$, weight decay of $10^{-5}$, and gradient clipping at 0.3. We selected the checkpoint with the lowest validation loss for evaluation. For environment transfer, the anechoic pre-trained models were fine-tuned for 20 epochs following~\cite{DeepEar}.

Input signals were framed using a 20~ms window with a 10~ms hop, and the ERB filterbank used $K=100$ bands. We evaluate five BiEAR variants for ablation: (1) \textbf{BiEAR w/o Control}, a passive model without the neural Q-controller; (2) \textbf{BiEAR + Single Controller + Abs}, which uses a shared controller for both ears with concatenated left/right subband SPLs as input and the absolute control strategy; (3) \textbf{BiEAR + Single Controller + Rel}, identical to (2) but using the relative control strategy; (4) \textbf{BiEAR + Dual Controller + Abs}; and (5) \textbf{BiEAR + Dual Controller + Rel}, both using dual controllers as described in Section~\ref{sec:method} with absolute and relative control, respectively. For all variants, the Q-factor was bounded to $[Q_{\min}, Q_{\max}]$ with $Q_{\min}=0.05$ and $Q_{\max}=30$, and the controller input was temporally smoothed with $\beta=0.8$. The absolute control used $\Delta Q_{\text{base}}=2$, $m_{\text{low}}=0.5$, and $m_{\text{high}}=5$, while the relative control used $\Delta Q_{\text{base}}=1$, $m_{\text{low}}=0.3$, and $m_{\text{high}}=5$, yielding a frequency dependent Q-variation range from 1.6 (lowest frequency) to 29.2 (highest frequency), following prior monaural adaptive front-end~\cite{buddhi_AdaFE}.

\vspace{-2mm}
\section{Results and Discussions}\label{sec:results}
\vspace{-0.5mm}
\subsection{Anechoic Environments}\label{sec:results-anechoic}
\vspace{-1mm}

Table~\ref{tab:performance} summarizes the performance of the baselines and BiEAR variants trained on \textit{Anechoic-train}, validated on \textit{Anechoic-val}, and evaluated on \textit{Anechoic-test} and the speaker disjoint set \textit{Anechoic-test-unseen-spk} under one-, two-, and three-speaker mixtures. We report sound detection accuracy, azimuth mean absolute error (MAE), and distance classification accuracy, computed as the mean over all eight azimuth sectors.
\begin{figure*}[!ht]
  \centering
  \vspace{-0.6em}
  \hspace*{-0.02\textwidth}%
  \includegraphics[width=0.69\linewidth]{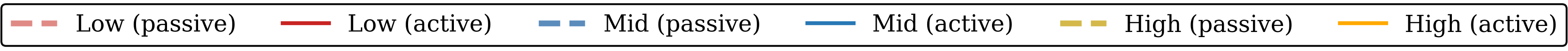}
  \vspace{-0.5em}
  \subfigure[Q variation over time and frequency.]{
    \includegraphics[width=0.32\textwidth]{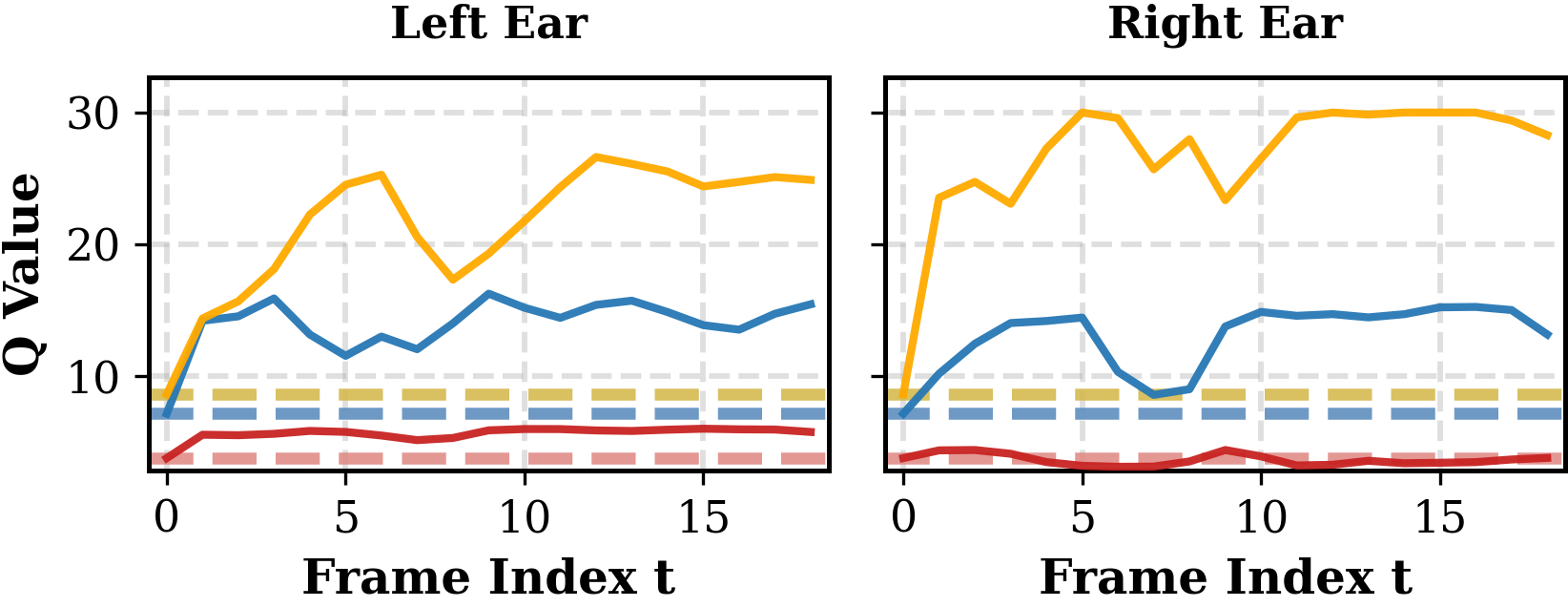}
    \label{fig:results_fig3_a}
  }
  \hfill
  \subfigure[Filter selectivity at a specific time frame.]{
    \includegraphics[width=0.32\textwidth]{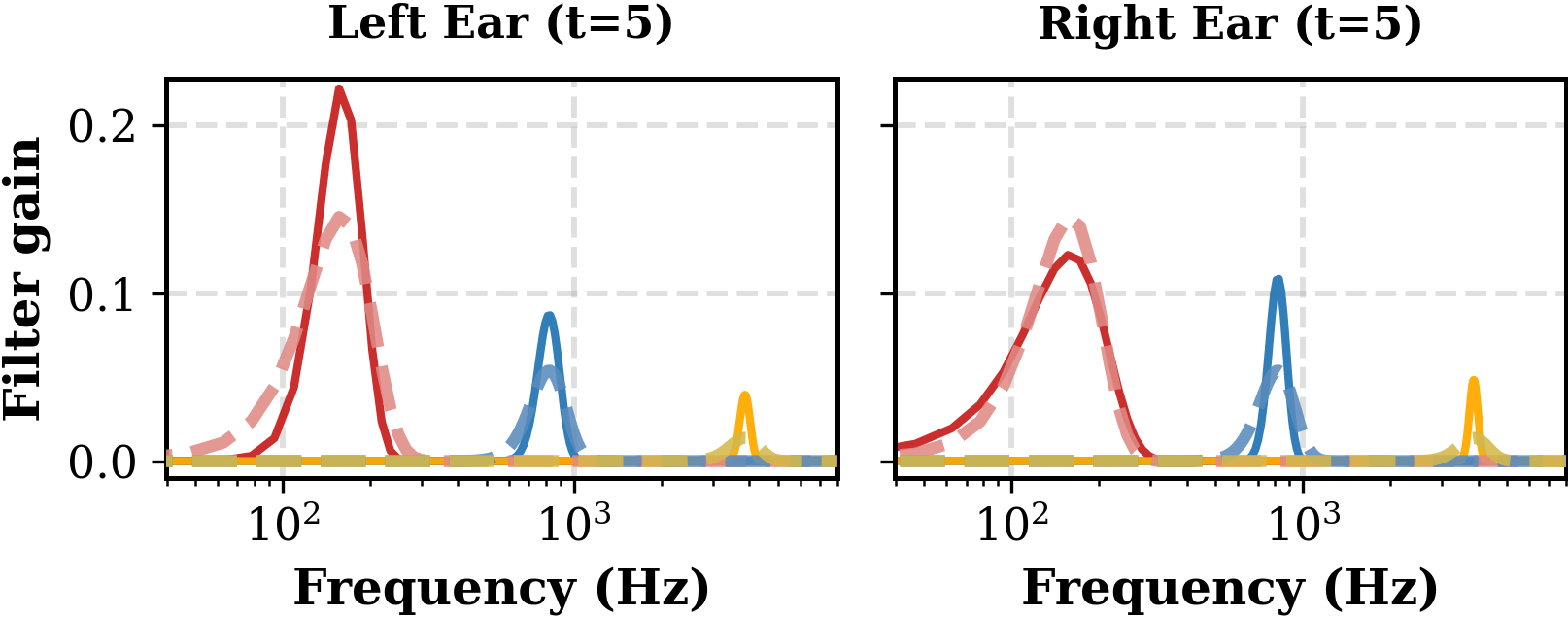}
    \label{fig:results_fig3_b}
  }
  \subfigure[Magnitude representation: passive vs. active.]{
    \includegraphics[width=0.32\textwidth]{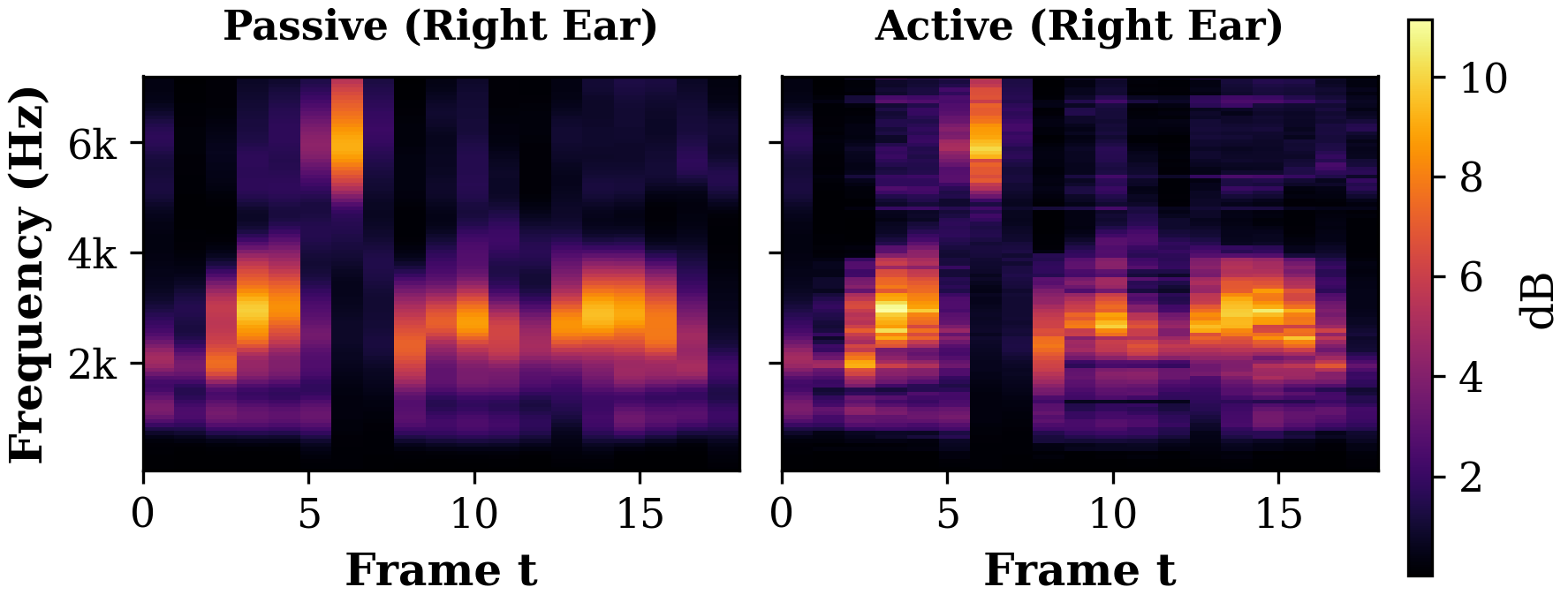}
    \label{fig:results_fig3_c}
  }
  \vspace{-1em}
  \caption{Adaptive filterbank behaviour of \textbf{BiEAR + Dual Controller + Rel.} for a single speaker at $292^\circ$ (front-right) and 2~m. Selected subbands: low $\approx$ 159~Hz, mid $\approx$ 821~Hz, and high $\approx$ 3.86~kHz. ``Passive'' is w/o controller; ``Active'' is adaptive.}
  \vspace{-4mm}
  \label{fig:results_fig3}
  \vspace{-2.5mm}
\end{figure*}

As shown in the results, incorporating the neural controller yields consistent gains across all speaker counts, and further confirms the advantage of the dual controller design over using a shared controller for both ears. We therefore adopt \textit{BiEAR + Dual Controller + Rel.} in subsequent experiments, as it provides the best overall performance. Compared with absolute control, relative control scales the modulation by the ERB-derived baseline $Q_0[k]$, resulting in more consistent subband bandwidth adjustments and improved training stability. BiEAR achieves comparable performance on seen and unseen speakers, motivating the use of unseen speaker test sets in the following diverse-room adaptation experiments. Compared with the baselines, the proposed active BiEAR improves sound detection and azimuth estimation, while AuralNet performs best on distance estimation, likely because its self-attention better emphasizes direct-path cues that benefit distance perception.
\vspace{-1em}
\subsection{Diverse Room Adaptation}\label{sec:results-reverb}
\vspace{-0.5em}
Table~\ref{tab:performance_rooms} reports results in two real rooms (meeting room and lecture hall) before and after environment transfer operation followed by the transfer learning approach in~\cite{DeepEar}, for the baselines and \textit{BiEAR + Dual Controller + Rel.} to evaluate adaptation to new acoustic conditions. Since Table~\ref{tab:performance} indicates that speaker identity has a negligible effect, the speakers used for room adaptation are all unseen. In the lecture hall, the recorded source distances do not match the discrete distance classes used for anechoic training (Table~\ref{tab:brir}), leading to generally lower distance estimation accuracy than in the meeting room. 

According to the results, DeepEar and AuralNet benefit substantially from environment transfer when adapting to unseen rooms, indicating a strong reliance on new environment adaptation for robustness. However, even after adaptation, their performance remains below that of the proposed model. In contrast, BiEAR achieves consistently strong results in both environments and is further improved more by environment transfer, demonstrating higher robustness and more effective adaptation. These findings support our design motivation: BiEAR can adapt rapidly and stably to reverberant conditions by learning the new room dynamics and adjusting its auditory selectivity tuning in a manner analogous to human auditory processing.
\vspace{-2.8mm}
\subsection{Visualisation and Analysis}\label{sec:results-visualisation}
\vspace{-1.65mm}
To better understand BiEAR, Fig.~\ref{fig:results_fig3_a} visualises the frame-wise Q-factor changes during inference for a single speaker binaural input, while Fig.~\ref{fig:results_fig3_b} shows how these variations change filter selectivity in representative low-, mid-, and high-frequency subbands at the $5^{\mathrm{th}}$ frame. In this example, the source is closer to the right ear. Accordingly, BiEAR increases right-ear selectivity (higher Q) in the mid/high bands, sharpening the response where head-shadow effects produce stronger and more reliable ILD cues. In the low band, BiEAR reduces the right-ear Q (wider bandwidth) while increasing the left-ear Q, supporting more phase samples in the right ear, consistent with time difference cues being most informative at low frequencies. Overall, the ear and time--frequency dependent Q modulation exhibits an interpretable, attention-like behaviour that emphasises the ear and subbands carrying the most informative spatial cues. As further shown in Fig.~\ref{fig:results_fig3_c}, active control also yields a more concentrated time--frequency magnitude distribution, enhancing salient regions and suppressing less informative ones, which supports robust back-end prediction.

\vspace{-3mm}
\section{Conclusion}
\vspace{-1.2mm}
In this work, we propose BiEAR, a human auditory-inspired adaptive binaural front-end for multi-speaker localisation and distance estimation. MOC inspired neural feedback regulates filterbank Q-factors, enabling time-frequency adaptive, ear-specific modulation. Across anechoic and real room conditions with unseen speakers and environments, BiEAR yields consistent gains in detection and azimuth estimation and remains robust to reverberation. Visualisations reveal interpretable, attention-like selectivity that emphasises informative subbands over time. The controller is an engineering abstraction of MOC efferent feedback, capturing the functional principle of feedback driven frequency selectivity rather than neural circuitry. Future work will extend BiEAR to additional binaural tasks and evaluate it on diverse datasets and moving sources.
\section{Acknowledgments}
This work was funded by ARC Discovery Grant DP210101228. The authors would also like to thank UNSW, Sydney, Australia, for providing PhD scholarship support.
\bibliographystyle{IEEEtran}
\bibliography{mybib}
\end{document}